\documentclass[Afour,sagev,times]{sagej}

\usepackage{moreverb,url}

\usepackage[colorlinks,bookmarksopen,bookmarksnumbered,citecolor=red,urlcolor=red]{hyperref}

\usepackage[utf8]{inputenc}
\usepackage[T1]{fontenc}
\usepackage{array}
\usepackage{booktabs}
\usepackage[table]{xcolor}
\usepackage[normalem]{ulem}

\newcommand\BibTeX{{\rmfamily B\kern-.05em \textsc{i\kern-.025em b}\kern-.08em
T\kern-.1667em\lower.7ex\hbox{E}\kern-.125emX}}

\begin{document}

\runninghead{Mazières and Roth}

\title{Large-scale diversity estimation through surname origin inference}

\author{Antoine Mazières\affilnum{1,2} and Camille Roth\affilnum{1,3}}

\affiliation{\affilnum{1}Centre Marc Bloch Berlin e.V., Computational Social Science team, Berlin, Germany\\
\affilnum{2}UMR-LISIS, INRA, Marne-la-Vallée, France\\
\affilnum{3}médialab, Sciences Po Paris, France}

\corrauth{Antoine Mazières, Centre Marc Bloch Berlin e.V.,
Friedrichstraße 191, 10117 Berlin, Germany}

\email{antoine.mazieres@gmail.com}

\begin{abstract}
The study of surnames as both linguistic and geographical markers of the past has proven valuable in several research fields spanning from biology and genetics to demography and social mobility. This article builds upon the existing literature to conceive and develop a surname origin classifier based on a data-driven typology. This enables us to explore a methodology to describe large-scale estimates of the relative diversity of social groups, especially when such data is scarcely available. We subsequently analyze the representativeness of surname origins for 15 socio-professional groups in France.

\end{abstract}


\maketitle

\newcommand{\tb}[1]{\textcolor{blue}{#1}}
\newcommand{\rk}[1]{\tb{\fbox{#1}}}

\section{Introduction}
\label{sec:zero}

Surnames have the objective property of designating a path in the ancestry tree, up to a point in time and space where the name was first coined and made hereditary.  While they are usually distant markers of an historical and geographical context, surnames still exhibit connections with present features and have thus been considered as a valuable proxy in population studies.
For one, surnames correlate with genetic proximity within populations \cite{jobling2001name,king2006genetic,lasker1985surnames} and have been diversely used to analyze human population biology \cite{lasker1980surnames}, identify cohorts of ethnic minority patients in bio-medical studies \cite{shah2010surname,polednak1993estimating,choi1993use}, improve research in genealogy \cite{king2009s} or describe the migration rates of human populations \cite{piazza1987migration}.
Social sciences more recently made use of surnames to statistically and indirectly appraise the composition of populations in various situations \cite{mateos2007review,mateos2014names}, including the demography of online\cite{chang2010epluribus,mislove2011understanding} and research\cite{wu2014science} communities, or the history of social mobility \cite{clark2014also,guell2012intergenerational}.

The purpose of the present article is twofold. First, it aims at assessing the possibility of building a general-purpose, worldwide surname origin classifier. Our approach combines elements which are already available in literature, and endeavors at enhancing both the learning data quality and broadening the geographical breadth and universality of surname origin typology. Second, we use this classifier to show that, despite its limitations at the individual level, it nonetheless enables simple and pertinent applications to the estimation of representation biases in origins in populations where no such data is explicitly available. We further illustrate its potential relevance for discrimination studies by comparing surname origin distributions for various sets of occupational groups and exam candidates in France.

\section{Statistically inferring a surname origin}

\subsection{Surname origin vs.~ethnicity}

Our approach relies essentially on the notion of surname \emph{origin} rather than ethnicity. 
Indeed, ethnicity is often defined \cite{weber1978economy,barth1998ethnic,tonkin2016history} as a \textit{subjective} feeling of membership to one or several groups or self-defined identities, composed of linguistic, national, regional and religious criteria. A quick glance at the present paper's bibliography reveals how much the academic literature aimed at inferring information from surnames relies on ethnicity to put names and individuals into groups, and derive subsequent analyses.

By contrast, a surname \emph{objectively} corresponds to a genealogical and traditionally patrilineal path whose origin coincides with the first appearance of this socially hereditary property in the family tree.  These moments vary much from one region to another, spanning from about 5,000 years ago in China to less than a century ago in Turkey.

Over 20 generations, the unique path of a name is one among more than a million (for about double the ancestors). Thus, in a randomly mating population, i.e.\@ without any kind of endogamy, this marker would assuredly carry extremely little information: given these figures, someone bearing a surname of a specific origin would not be more likely to exhibit characteristics found in other bearers of a surname of the same origin. However, the existence of a strong endogamy among humans --albeit probably decreasing \cite{rosenfeld2008racial}-- entails a correlation between surnames and the preferences that characterize this endogamy: geographical proximity, social and economical status, languages, political, genetic, regional and religious criterias. Put simply, as a result of, say, geographical endogamy, the correlation between the geographical origins of the father and the mother of a person induces a correlation between the geographical origins of their surnames, whereby the father name partly informs on the geographical origin of the mother. This phenomenon is likely the common cause behind the significance of the results found in the above-cited studies.

With this in mind, ethnicity appears as a potentially uncertain detour through a context-dependent and highly subjective matter, while the reference to an origin offers a more objective description of the variations in features extracted from surnames. To speak of origins nonetheless demands that we make a decision on how we partition the world into distinct regions.  At the very low level, to make matters simple and comparable, we first decided to use the present-day list of countries, acknowledging that no spatial or temporal partition of the world would be likely to take into account the wide diversity of overlaps between territories and populations at various points in time. 

\subsection{Crafting the learning data}

\begin{figure}[!htb]
    \includegraphics[height=22.6cm]{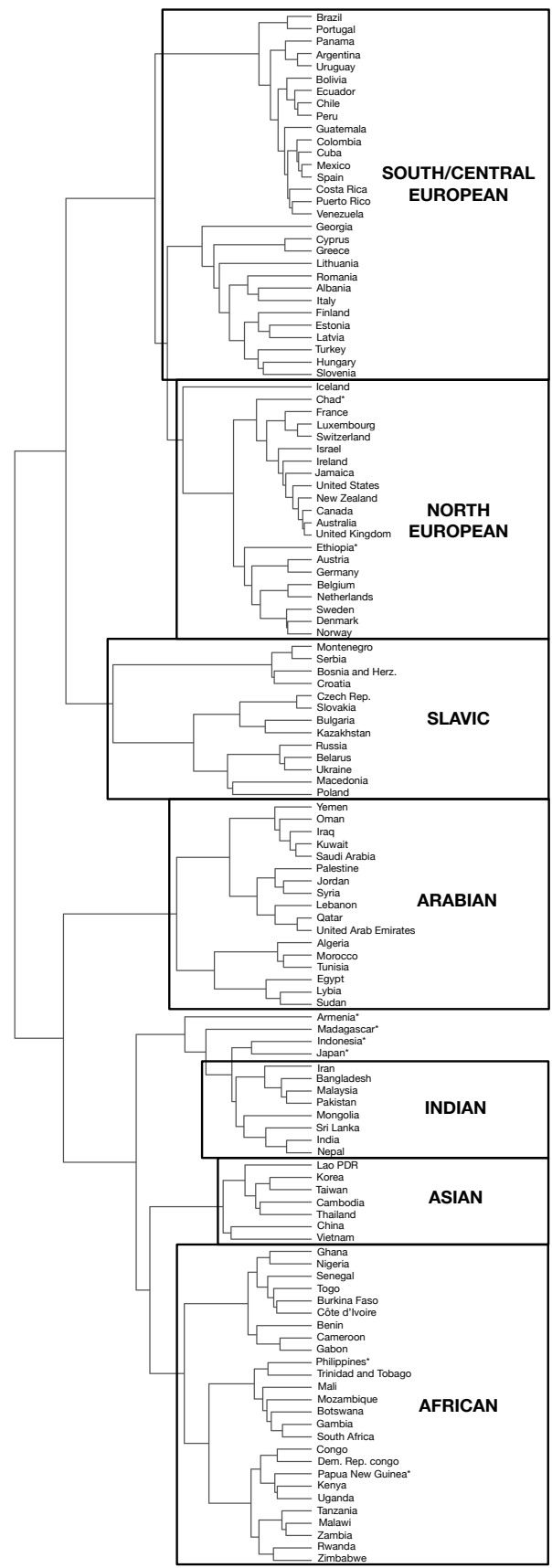}
    \caption{Clusters of surname origins\\
\footnotesize{\textit{Countries marked by a star (*) are interpreted as misclassified and reassigned in the following manner: Philippines, Japan and Indonesia are assigned to the Asian cluster, Ethiopia to African. Papua New Guinea, Madagascar, Jamaica, Chad and Armenia are deleted from the dataset as they represented a very low number of initial observations.}}}
    \label{hclust}
\end{figure}

How could we, humans, be able to form an intuition on the origin of some surnames ? If one has never encountered the name ``\textit{Toriyama}'', one might still correctly make a guess on its Japanese origin, for instance because of the way it sounds when being pronounced, or the pattern of letter ordering. This admittedly hints at the existence of a second, closely-related proxy: surnames were originally coined (and have also been modified) by speakers belonging to a given linguistic space. Some structural and recurrent linguistic properties are more likely to be found in surnames of the same origin.  

Thus, we aim at creating a classifier able to infer sufficiently well the probable origin of a  surname from its spelling. To take a simple example, the distribution of letters in a text usually yields a good prediction of its language, assuming sufficiently many words and prior knowledge of empirical distributions for a set of languages.  While it would be ambitious to expect a decent precision from surname single letter distributions, the use of \emph{subsets} of letters, including morphemes, appears much more promising.  To define \emph{learning features}, we thus decompose all surnames into various subsets of letters of size $n$, or ``n-grams''. This eventually constitutes the feature set for the whole dataset. We then describe a given surname by its distribution on these features. 

Building a statistical model able to reproduce the above intuition at large scale for all origins means that we must first fit the model by using a large and diversified number of surnames labeled with their origins, or \emph{training dataset}.
To gather such learning examples, previous works relied on a variety of explicitly labeled sources including census data \cite{mislove2011understanding}, Olympic game participant records \cite{leename}, phone books \cite{mateos2014names} or even Wikipedia data \cite{ambekar2009name}. 

Another study used the PubMed search engine to extract scientific bibliographical records\cite{torvik2016ethnea}. We follow a similar approach since this open data source\endnote{Using the query \texttt{1800:2020[dp]} on \url{https://www.ncbi.nlm.nih.gov/pubmed/}} enables easy reproductibility of our research and provides an extensive volume of references with more than 25 million publications. For each record, we extracted author surnames and their affiliations when they were related to one of the 176 countries of the Natural Earth dataset\endnote{Natural Earth Data, 1:110m Cultural Vectors, \url{http://www.naturalearthdata.com/downloads/110m-cultural-vectors/}}. 

We assume that surnames whose affiliation distribution is heavily peaked for a given country are more likely to originate from that country. However, using PubMed data suffers from several biases, among which:
\begin{itemize}
	\item The increased nomadism of the scientific population, lowering the quality of the affiliation as a reliable origin.
	\item The heterogeneous academic activity of countries, over-sampling the most productive ones at the expense of others.
	\item The potential bias of medical publication databases in favor of Anglo-Saxon publication venues\cite{Nieminen:aa}, under-sampling the rest of the world.
\end{itemize}

A first obvious step for counterbalancing these biases consists in considering surname frequencies, \hbox{i.e.} normalizing surname occurrences in a given country by the total number of occurrences for that country. Then, in an effort to restrain our training dataset to true positives, we use a measure of statistical dispersion, the Herfindahl–Hirschman Index (HHI)\cite{herfindahl1950concentration,hirschman1980national}, to identify names whose presence is highly concentrated in one country only. We request a HHI of at least 0.8 as well as a maximal frequency over all countries of at least 0.0001 \%.
Even though this method eliminates some of the most common names, for they are susceptible to have spread all over the world, it narrows our focus to a set of about 650k surnames which we call ``core names'' and which we assign to the country where frequency is maximal.

\subsection{A data-driven typology of surname origins}

Nonetheless, the number of these core names remains unevenly distributed across countries, partly as a result of the above-mentioned under-sampling. It goes from 163 names for Montenegro to 41k names for Spain, with an overall average of 5\,145. Before training our model, we thus need to introduce coarser categories to achieve a minimal significance for each geographic area. 

Keeping in mind the eventual goal of appraising over- and under-representation of origins in socio-professional groups, we conservatively decide to categorize countries into a relatively small number of world regions.  To do so, we first cluster countries according to the training features.
More precisely, we created a large ``country / n-gram'' matrix whose rows are countries and columns are n-grams of core names: a cell indicates the frequency of a given n-gram among the core names of a given country. We then performed hierarchical clustering on this matrix using Ward linkage\cite{ward1963hierarchical}. This yields the dendrogram shown in Figure \ref{hclust} from which we may extract 7 rough categories of surname origins. We concretely aggregate countries by following the dendrogram in a monotonous manner from the bottom to the top while avoiding to merge categories belonging to strongly unrelated geographical areas. This process creates what appears to be an interpretable regionalization of the world at the cost of a very limited number of inconsistencies.

We relabel the original ``surname-country'' associations according to these clusters. We eventually train a classifier on this new ``surname-world region'' dataset, using the same learning features. Broadly, a classifier is a model (and, in practice, a function) which takes as inputs the learning features for a given observation (in our case, a surname and its letter subsets) and outputs a guessed label (in our case, an origin under the form of a world region).

The state of the art features a variety of methods such as hidden markov models and decision trees \cite{ambekar2009name}, recurrent neural networks \cite{leename} or logistic regressions \cite{torvik2016ethnea}. We focus on one of the most classical classifiers, called Naive Bayes, which in our case yielded the best overall results among a variety of other traditional approaches available in Scikit-learn\cite{pedregosa2011scikit}, the python classification algorithm library we used.\endnote{We concretely apply a multinomial naive Bayes model with an additive (Laplace/Lidstone) smoothing parameter of 0.1. A programming notebook is available to observe and reproduce all steps described here: \url{https://namograph.antonomase.fr/}}
Naive Bayes is a simple classifying technique consisting in estimating the probability that an object belongs to a certain class given a set of observed features. It applies the Bayes theorem on the probabilities that surnames exhibit certain features knowing that they belong to some origin. It additionally relies on the assumption that these features are statistically independent, \hbox{i.e.} the contributions of each of these features to the target probability are independent from one another, hence the ``naive'' qualification. In practice, we train the model on about 85\% of the core name dataset while keeping aside about 15\% of the core name dataset to evaluate model performance.

\begin{table}
\footnotesize\sf
\caption{\label{tab:corenames}Number of core names (totals, while around 15\% are used for the evaluation) and classifier performances for each cluster in terms of precision and recall.}
\centering
\begin{tabular}{lrrrr}\toprule
\textbf{Cluster}              &  \multicolumn{2}{c}{\textbf{Core names}} &  \multicolumn{2}{c}{\textbf{Class. Perf.}}\\
&  \emph{Total} & \emph{Evaluation} &  \emph{Precision} & \emph{Recall} \\\midrule
African	             & 30\,748 & 4\,529 & 0.43 & 0.61 \\
Arabian	             & 31\,272 & 4\,596 & 0.52 & 0.72 \\
Asian	             & 44\,658 & 6\,754 & 0.61 & 0.77 \\
CS-European &189\,624 &28\,668 & 0.81 & 0.71 \\
Indian	             & 68\,145 &10\,067 & 0.63 & 0.72 \\
N-European	     &216\,465 &32\,469 & 0.78 & 0.62 \\
Slavic	             & 65\,259 & 9\,843 & 0.64 & 0.84 \\
\bottomrule\em Total                     & \em 646\,171& \em 96\,926 & & 
\end{tabular}
\end{table}

Classification performance is shown in Table \ref{tab:corenames} and is expressed in terms of \emph{precision} and \emph{recall}, along with the corresponding set sizes. For instance, the model achieves a precision of 61\% for Asian and a recall of 77\%, meaning that 61\% of names guessed as ``Asian'' belong to the Asian cluster, while 77\% of names belonging to the Asian cluster are correctly guessed (recalled by the model) as ``Asian''. Success differs significantly from one class to another, with very satisfying results for the Central/South European and Slavic clusters and quite moderate performance for the African cluster. 
How much of this error is due to the lack of academic data in certain areas or the difficulty to identify pattern in surnames of a specific area is yet to be determined.

Notwithstanding, since we are interested in comparing the over- and under-representation of surname origins \emph{between} two socio-professional populations of a given country, we contend that this type of error does not significantly jeopardize our aim.  We first postulate that classification errors for a given surname origin remain homogeneous from one dataset to the other, i.e. that the names of a given origin are globally going to be classified (and misclassified) with the same success in both datasets. In other words, irrespective of their proportion within a given dataset, we assume that all surnames of, say, Indian origin, will be as often correctly recalled by our algorithm as Indian in all datasets, i.e. 71.8\% of the time (and errors will be distributed across other origins in similar proportions for all datasets). Put differently, we suppose that names which pose inference problems \hbox{w.r.t.} our model are roughly distributed homogeneously and are not biased across datasets (for example, if ``Toriyama'' is misclassified, we assume that it is no more or less present among Asian names in one dataset than in another one).

We nonetheless have to consider that classification errors vary across origins. This is shown by the confusion matrix on Table~\ref{tab:recall}.  Here, names of Arabian origin are guessed as Asian 2.46\% of the time, while it is about 7.04\% for names of African origin.  Even if the above assumption enables us to use the same confusion matrix for all datasets, we still have to adjust guesses knowing that the algorithm exhibits some propension to over-/under-estimate depending on the origin. In other words, knowing that a proportion of names which \emph{actually} belong to a given origin are \emph{guessed} as belonging to another origin, we correct guesses to infer back the probability of actual origin for a given guess $P(\text{actual}=j | \text{guessed}=i)$. In practice, we multiply guessed numbers of surname origins by this probability which we extract from $C$ by Bayesian inference.\endnote{More precisely, we compute $P(\text{actual}=i | \text{guessed}=j)$ as $C_{ij}/\sum_{j'}C_{ij'}$. Moreover, since the confusion matrix is computed using prior proportions of surname origins extracted from Pubmed, it is likely to be based on priors which very significantly diverge from the average proportions of surname origins in the ``general'' French population. To accommodate for the Pubmed bias as much as possible, we adjust the priors of the confusion matrix so that they match a distribution guessed initially by the uncorrected classifier on the \texttt{Brevet} dataset. This uncorrected distribution yields respectively 4.8, 8.3, 3.1, 20.7, 3.4, 57.1 and 2.6 \% for each of the origins: African, Arabian, Asian, Central SE, Indian, NE, and Slavic. We thus correct the original confusion matrix of Table~\ref{tab:recall} by making column sample sizes proportional to these figures. In other words, the confusion matrix that we eventually use exhibits a structure more similar to that of the initially guessed Brevet proportions than the Pubmed ones.}

\newcommand{\fz}[1]{\emph{\scriptsize #1}}
\newcommand{\pfz}{\multicolumn{1}{r|}{\fz{\tiny \%}}}
\begin{table}
\footnotesize\sf
\caption{\label{tab:recall}Confusion matrix $C$. This matrix shows the number of names from the evaluation sets (see Tab.~\ref{tab:corenames}) of an actual origin (in columns) which are guessed as belonging to a given origin (in rows). The first subrow indicates total numbers, the second subrow refers to proportions within an actual origin.}
\centering
\resizebox{\columnwidth}{!}{%
\begin{tabular}{l|ccccccc}\toprule
\textbf{Guessed}&\multicolumn{7}{c}{\textbf{Actual origin}}\\
\textbf{origin}& Afr. & Arab. & Asian & CSE & Indian & NE &Slavic\\\cmidrule{1-8}
African
&\textbf{2763} & 165 & 381 & 1081 & 460 & 1441 & 157 \\
\pfz{}&\fz{\textbf{61.0}} &   \fz{3.59}        &   \fz{5.64}        &   \fz{3.77}        &   \fz{4.60}    &       \fz{4.44}    &   \fz{1.60} \\
\rowcolor{gray!15}
Arabian
&159	&\textbf{3292}	&84	&577	&598	&1549	&77	\\
\rowcolor{gray!15}
\pfz{}&\fz{3.51}          &\fz{\textbf{71.6}} &  \fz{1.24}        &   \fz{2.01}        &   \fz{5.94}    &       \fz{4.77}    &   \fz{0.78} \\

Asian
&319	&113	&\textbf{5200}	&831	&716	&1147	&174	\\
\pfz{}&\fz{7.04}          &   \fz{2.46}        &\fz{\textbf{77.0}}  &   \fz{2.90}        &   \fz{7.11}    &       \fz{3.53}    &   \fz{1.77} \\
\rowcolor{gray!15}
CS-Eur.
&258	&128	&274	&\textbf{20364}	&299	&3535	&324	\\
\rowcolor{gray!15}
\pfz{}&\fz{5.70}          &   \fz{2.79}        &   \fz{4.06}        &\fz{\textbf{71.0}}  &   \fz{2.97}    &       \fz{10.9}    &   \fz{3.29} \\
Indian
&273	&487	&420	&991	&\textbf{7226}	&1862	&191	\\
\pfz{}&\fz{6.03}          &   \fz{10.6}        &   \fz{6.22}        &   \fz{3.46}        &\fz{\textbf{71.8}}&     \fz{5.73}    &   \fz{1.94} \\
\rowcolor{gray!15}
N-Eur.
&643	&351	&315	&3254	&609	&\textbf{20183}	&670	\\
\rowcolor{gray!15}
\pfz{}&\fz{14.2}          &   \fz{7.64}        &   \fz{4.66}        &  \fz{11.4}         &   \fz{6.05}    &\fz{\textbf{62.2}}  &   \fz{6.81} \\
Slavic
&114	&60	&80	&1570	&159	&2752	&\textbf{8250} \\
\pfz{}& \fz{2.52}         &   \fz{1.31}        &   \fz{1.18}        &  \fz{5.48}         & \fz{1.58}      &\fz{8.48}&\fz{\textbf{83.8}}\\
\bottomrule
&4529 & 4596 & 6754 & 28668 & 10067 & 32469 & 9843\\
\pfz{}&\fz{100}&\fz{100}&\fz{100}&\fz{100}&\fz{100}&\fz{100}&\fz{100}
\end{tabular}}
\end{table}

\section{Estimating origin-based discrimination in France}
\label{sec:two}

\newcommand{\ttt}[1]{\emph{#1}}

\subsection{Datasets and estimation methodology}

\begin{figure*}[!ht]
\begin{center}
\includegraphics[width=.92\linewidth]{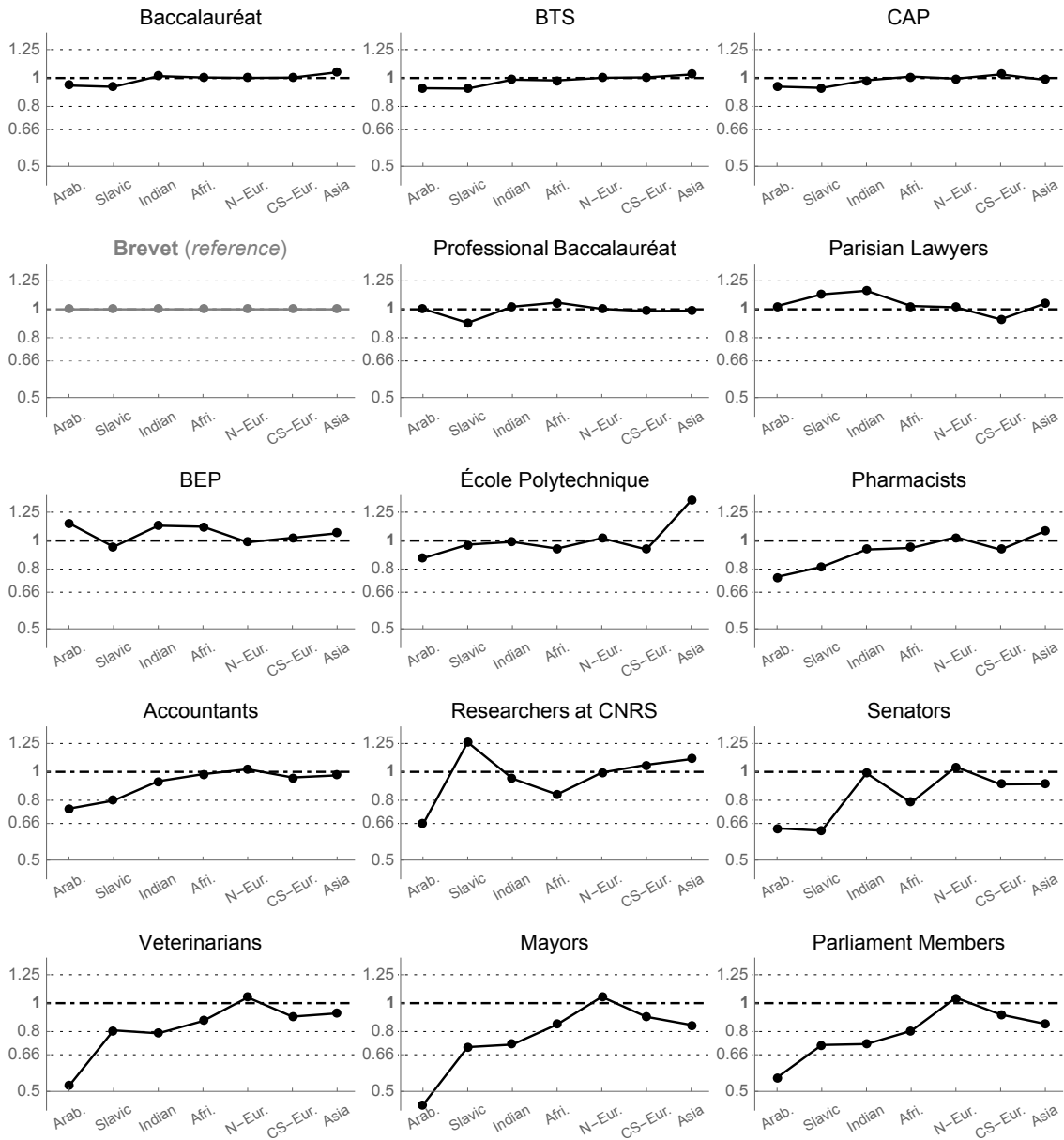}
\end{center}
\caption{\label{reprs}Over-/under-representation of surname origins among all datasets. Each graph shows the ratios between the target dataset and the reference dataset (\ttt{Brevet}) for each origin category.  A logarithmic scale is used to depict equivalent over- or under-representation ratios at equal distance from the y=1 reference line.}
\end{figure*}

We now illustrate the method on 15 datasets representing various areas of French society, see Table~\ref{datasets}. Three datasets are linked to political functions (\ttt{Mayors}, \ttt{Parliament Members} and \ttt{Senators}), five of them represent various types of occupations (\ttt{Pharmacists}, \ttt{Lawyers}, \ttt{Accountants}, \ttt{Veterinarians}, \ttt{Researchers}), and six are made of lists of candidates to various state exams (\ttt{Brevet}, \ttt{Baccalauréat}, \ttt{BEP}, \ttt{CAP}, \ttt{BTS}, \ttt{Professional Baccalauréat}). The \ttt{École Polytechnique} dataset lists students at one of the most highly-ranked engineering school in France. 

From the list of surnames of each dataset, we apply the classifier to obtain vectors of values representing the guessed distributions of surname origins according to our typology. Note that this approach works by construction at the level of groups and may not be used at the level of individuals: to take an example from a distinct context, if we know that the given name ``Camille'' is about 80\% of the time a female name, we are not able to draw a precise conclusion on the gender of a given Camille, while we can say that a group of 100 Camille is likely to be around 80\% female.

\rowcolors{2}{gray!15}{white}
\begin{table}[h]
\scriptsize\sf
\caption{List of datasets along with the corresponding numbers of observations.\label{datasets}}
\begin{tabular}{p{2.3cm}p{4cm}r}

\toprule
Name&List of surnames of all \ldots&nb. obs.\\
\midrule
\ttt{Brevet}& Candidates to \textit{Diplôme National du Brevet} in 2008 \endnote{Source for all 2008 exams: \url{http://www.bankexam.fr/resultat/2008}}&562,952\\
\ttt{Baccalauréat}& Candidates to the nationwide \textit{Baccalauréat} (\textit{Général} and \textit{Technologique}) in 2008&435,645\\
\ttt{BEP}& Candidates to \textit{Brevets d'Études Professionnelles} in 2008&116,814\\
\ttt{CAP}& Candidates to \textit{Certificats d'Aptitude Professionnelle} in 2008&98,364\\
\ttt{BTS}& Candidates to \textit{Brevets de Technicien Supérieur} in 2008&87,917\\
\ttt{Professional Baccalauréat}& Candidates to \textit{Baccalauréats Professionnels} in 2008&80,672\\
\ttt{Pharmacists}& Pharmacists registered in their \textit{Ordre Professionnel} in 2017 \endnote{Source: Online directory of the \textit{Ordre National des Pharmaciens}, \url{http://www.ordre.pharmacien.fr/annuaire/pharmacien}}&73,422\\
\ttt{Mayors}& Mayors of French cities (``communes'') in 2014 \endnote{Source: French gouvernment open data repository, \url{https://www.data.gouv.fr/fr/datasets/liste-des-maires-au-17-juin-2014/}}&36,628\\
\ttt{Parisian Lawyers}& Lawyers registered in the Parisian Bar Association in 2017 \endnote{Source: Online directory of the Parisian Bar Association, \url{http://www.avocatparis.org/annuaire}}&32,021\\
\ttt{\'Ecole Polytechnique}& Students at \textit{École Polytechnique} (1958-2016) \endnote{Source: Alumni online directory of École Polytechnique, \url{https://www.polytechnique.org/search}}&23,058\\
\ttt{Accountants}& Accountants registered in their \textit{Ordre Professionnel} in 2017 \endnote{Only independent, salaried and honorary accountants. Source: Online directory of the \textit{Ordre National des Experts-Comptables}, \url{http://www.experts-comptables.fr/annuaire}}&20,946\\
\ttt{Veterinarians}& Veterinary physicians registered in their \textit{Ordre Professionnel} in 2017 \endnote{Source: Online directory of the \textit{Ordre National des Vétérinaires}, \url{https://www.veterinaire.fr/outils-et-services/trouver-un-veterinaire.html}}&15,710\\
\ttt{Researchers}& Researchers at \textit{Centre National de la Recherche Scientifique} in 2017 \endnote{Only tenured researchers. Source: CNRS Online directory, \url{https://annuaire.cnrs.fr/l3c/owa/annuaire.recherche/index.html}}&12,657\\
\ttt{Parliament Members}& Parliament Members of \textit{Assemblée Nationale} (1958-2016) \endnote{Source: French National Assembly online databasem \url{http://www.assemblee-nationale.fr/sycomore/liste_legislature.asp?legislature=48}}&8,326
\end{tabular}\\[10pt]
\end{table}

In order to show how the diversity in terms of surname origins of certain subgroups of the population departs from that of a common reference point, we rather focus on dataset-to-dataset comparisons rather than raw distributions.
In other words, comparing surname origin distribution across datasets enables us to assess the extent and magnitude of the divergence in the representativeness of groups of people with a given surname origin and, more broadly, the fact that some datasets and some origins exhibit the same pattern of divergence, likely indicative of similar underlying processes.

There is no public and unbiased source of data which covers surnames of the  French population in order to perform such comparaisons. Therefore, we chose the \ttt{Brevet} dataset as a point of comparison since it represents the most widely passed exam in France and therefore, a wide sample of people who lived in France and were generally aged 14-15 as of 2008, hence 23-24 as of 2017. As such, it is also likely to exhibit a bias towards younger people.  

Simply calculating the ratio between each target dataset and \ttt{Brevet} yields the results shown in figure~\ref{reprs}, which enables the observation of several profiles of representativeness among the datasets described in table \ref{datasets}. As such, values higher (\hbox{resp.} lower) than 1 correspond to surname origins which are over-represented (\hbox{resp.} under-represented) compared with their presence in \ttt{Brevet} (logically, \ttt{Brevet} exhibits a flat profile where all origins have a ratio of 1).  Of course, these ratios do not render the fact that some categories are significantly more populated than others: this is typically the case for ``North European'', which is the most common surname origin found in these French datasets. As a result, large under- or over-representation of less populated categories may have a relatively marginal effect on the over- or under-representation of the most populated category. The graphs of figure~\ref{reprs} should be read with this provision in mind: because of their sheer presence in all datasets, North European surnames ratio are often grouped around 1, while other categories may vary significantly below or above 1.  In other words, these ratios tend to emphasize the over- or under-representation of minority categories, rather than the strong presence of the majority category --- this can prove useful in the context of discrimination studies.

Besides, datasets and origins can be grouped according to the similarity of their divergence profiles, using for instance a simple hierarchical clustering based on the Canberra distance. Graphs in figure~\ref{reprs} have been organized according to this proximity in order to display and emphasize datasets or origins behaving in a comparable manner.

\subsection{Preliminary results: observations}

While we do not aim at discussing in detail the implications of such and such bias in some dataset, we may emphasize a few trends to illustrate the interpretation of the results. 

All elective political functions (\ttt{Mayors}, \ttt{Parliament Members}, \ttt{Senators}) together with \ttt{Veterinarians}, exhibit a marked over-representation of Northern European surnames. On the other hand under-representation, when it appears in these four datasets, is much more pregnant than in other datasets. It is actually spread among the remaining origins by following a comparable pattern across all four datasets, with Arabian names being the most significantly affected, closely followed by Slavic, Indian and African surnames.

State exams show an overall smoother profile, with \ttt{Baccalauréat}, \ttt{BTS} and \ttt{CAP} having the almost exact same distributions while \ttt{Professional Baccalauréat} and \ttt{BEP} display slightly different configurations. 
Interestingly, some datasets exhibit specific over-representation peaks for a single surname origin, such as Asian for \ttt{École Polytechnique} and Slavic for \ttt{Researchers}.

Some additional patterns emerge by examining these results along origins rather than datasets. For one, the under-representation of Arabian names is constant across all datasets, to the exception of \ttt{BEP}. Surnames of Asian origin are generally under-represented in elective functions, while their strongest over-representation occurs for two groups related to higher education, \ttt{École Polytechnique} and \ttt{Researchers}. As said above, North European surnames represent, in absolute numbers, the bulk of inferred origins. Their representativeness is generally close to 1, indicating no remarkable \emph{relative} variation across datasets, from \emph{Brevet} to \emph{Parliament Members}, even though the ratio rises slightly above 1 for the last four datasets, possibly as a result of the strong under-representation of other origins.

\subsection{Contribution scope and future work}

There is an undefined leap between the statistical observation of representativeness and fairness, between under-representation and discrimination and between over-representation and privilege. For instance, while we can say that a discrimination often implies an under-representation, the inverse is not necessarily true\cite{jobard2007color} and discrimination is usually evaluated on multiple complementary dimensions\cite{delattre2013introduction}, both qualitative and quantitative. 
Moreover, our method does not yet take into account other socio-demographic variables to control for the existence of common causes to the results. This would make it possible to say that, for instance, there is more of such surname origin in a given dataset because such origin is over-represented in such socio-demographic segment, which is itself over-represented in the population of the said dataset. Taking for example the most under-represented case in our results --- mayors with Arabian surnames --- one may conclude that it illustrates a well-known discrimination in France\cite{foroni2016discrimination,cediey2007discriminations} towards people of Arabian origins in elective functions. However, it is unclear how much of this ratio may be explained by discrimination or, for instance, by a uneven geographical presence of a given group of immigrants, or descendants thereof \cite{brutel2016localisation}.

The results presented here should be further examined and perhaps challenged both for their statistical significance and historical relevance. In this respect, our article simply acknowledges that the study of representativeness is part of discrimination studies, whereby methods for large-scale estimation of the former contribute to the latter. Our results could be nuanced with expertise of the demography of each socio-professional group considered here together with in-depth knowledge of the history of colonization and immigration in France\cite{beauchemin2016trajectoires}.

Finally, while we applied our methodology to datasets representative of certain groups in French society, comparisons with other contexts (countries, world regions, transnational entities) with the help of relevant surname datasets could yield fruitful insights on both our method and on possible interpretations.

\subsection{Concluding remarks}

The aim of this article lies in demonstrating the feasibility of a technique of estimation of representativeness based on a combination of open data sources, in contexts where data explicitly documenting individual origins may be difficult to process. We endeavored at showing that these methods can work in the absence of public data and/or data specifying distribution priors\cite{chang2010epluribus} or a priori ethnic taxonomies\cite{ambekar2009name}.

By making the model available to anyone and relying on data open sources, we hope to encourage further exploration and improvements of such techniques, especially in the context of discrimination studies and the discussion of the specific biases corresponding to present and future datasets.

\begin{acks}
The authors would like to thank Telmo Menezes, Mikaela Keller, Elian Carsenat, Jean-Philippe Cointet, Élise Marsicano, Fabien Jobard, Jérémy Levy and Mélanie Bourgeois for their help with this research.
\end{acks}

\begin{funding}
This paper has been partially supported by the ``Algodiv'' grant (ANR-15-CE38-0001) funded by the ANR (French National Agency of Research).
\end{funding}

\theendnotes

\bibliographystyle{SageH}
\bibliography{biblio}
\end{document}